\documentclass[11pt]{article}  
\usepackage{cite}
\usepackage{epsfig}
\usepackage{amsmath,amssymb}
\def\shiftleft#1{#1\llap{#1\hskip 0.04em}}
\def\shiftdown#1{#1\llap{\lower.04ex\hbox{#1}}}
\def\thick#1{\shiftdown{\shiftleft{#1}}}
\def\b#1{\thick{\hbox{$#1$}}}
\begin{document}
\title{A possible quark origin of two-pion emission
in $e+N$ and $N+N$ collisions}%
\author{I.T. Obukhovsky\footnote{The talk at International 
Symposium MENU2001, the George Washington University, 26-31 July.}\\
Institute of Nuclear Physics, Moscow State University,
Moscow 119899, Russia}%
\maketitle
\begin{abstract}
The constituent quark model has been taken as a starting point
for investigation of one- and two-pion emissions. A wide range of
observables  have been evaluated in the framework of the $^3P_0$
model. Comparison of the results with the data (when it is
possible) shows an approximate agreement.
\end{abstract}
\section{Introduction}

In recent works of the Moscow \cite{n} and Moscow-Tuebingen
\cite{k} groups the $^3P_0$ model of pion emission \cite{f,d} has
been extended to the description of mesonic clouds in
nucleonic systems. In this model the meson momentum distribution
in the cloud replicates the quark momentum distribution in the
nucleon or in the overlap region of a two-nucleon system, and thus
the standard quark-shell-model technique can be used for evaluation
of meson cloud contributions to varied processes.
As a result the pion emission amplitudes of interest [e.g.
the amplitudes $M(N\to\pi+B)$ of virtual transitions $N\to B+\pi$
for pion quasielastic knockout $N+e\to e^{\prime}+\pi+B$ with
$B=N,\,\,\Delta,\,\,N_{1/2^+}(1440)$, etc.] have been calculated
by this technique without invoking the Lagrangian field theory
coupling constants $f_{\pi NB}$ and vertex form factors $F_{\pi
NB}(k^2)$ (mostly unknown). On the same footing the related
two-nucleon vertices $N+N\to ^2B+\pi$, actual for quasielastic
processes of a similar type (e.g. $d+e\to e^{\prime}+\pi+^2B$),
have been evaluated.

\noindent
 In reactions $N+e\to e^{\prime}+\pi+B_i$ ($d+e\to
e^{\prime}+\pi+^2B$) a pionic decay of the final baryon (dibaryon)
leads to a second pion emission $B_i\to N+\pi$ ($^2B\to N+N+\pi$)
with a momentum distribution predicted by the $^3P_0$ model as
well. However, the predicted cross sections for respective
coincidence experiments are too small~\cite{n} to be seen over the
background. Nevertheless another two-pion processes, e.g. the
virtual "$\sigma$-meson" (correlated $\pi\pi$ pair) emission, have
been considered.

\section{One-pion predictions}

Table 1 shows the "spectroscopic factors" \cite{n} $S_N^{\pi
B}=\int\,\overline{|\Psi_N^{\pi B}(\bf k)|^2}\,d^3k$ for pion
momentum distributions $\Psi_N^{\pi B_i}(\bf k)$ in varied
channels $N\to \pi+B_i$, $B_i=N$, $\Delta$, $N^*$, etc.
\[
 \Psi_p^{\pi B_i}({\bf k})\sim
 \,\, \frac{M(N\to\pi+B_i)}
 {E_N({\bf k})-M_N+\omega_{\pi}({\bf k})} ,\quad
 \overline{|M(N\to\pi+B_i)|^2}=f_{\pi qq}^2\,
 \frac{4M_NM_B}{m_{\pi}^2} C_{\pi NB_i}^2{\bf k}^2\,
 F_{\pi NB_i}({\bf k}^2),
\]
 with coefficients
 $C_{\pi NB_i}^2$ calculated algebraically in the framework of
 $^3P_0$ model ($f_{\pi qq}=3/5\,f_{\pi NN}$ is the effective
 $\pi q\bar q$ coupling constant common for all the channels).

\medskip

 \centerline{
{\bf Table 1.} Spectroscopic factors $S_p^{B_i\pi}$ and pion decay
widths $\Gamma_{B_i}$ for some baryons.}

 \centerline{
 \begin{tabular}{c|c|c|c|c}
 $B_i$&$N$&$\Delta$&$N_{1/2^-}(1535)$&$N_{1/2^+}(1440)$\\[4pt]
 \hline
 \hline
 $C_{\pi NB_i}^2$&$2$&$\frac{64}{25}$&
 $\frac{16\omega_{\pi}({\bf k}^2)}{75{\bf k}^2}$&$\frac{2}{27}$\\[6pt]
 $\Gamma_{B_i}$(MeV)&-&63&52&47\\[6pt]
 $S_p^{B_i\pi}$&0.153&0.065&0.006&0.011
   \end{tabular}}

\medskip

\noindent
 The calculated values $\Gamma_{B_i}$ are realistic ones but they
are no more than in a rough agreement with experimental data.
Nevertheless the evaluations made on the basis of $^3P_0$ model
have some advantages over the standard phenomenological
description. In the model all the form factors $F_{\pi NB_i}({\bf
k}^2)$ depend only on two parameters: $b\quad\mbox{and}\quad
b_{\pi}/b$, where
 $b\approx 0.6$ fm and $b_{\pi}\approx 0.3$ fm are standard free
 parameters of the constituent quark model (the baryon and pion
radius correspondingly). It is interesting that the sum of all the
calculated spectroscopic factors $N_{\pi}=\sum_{i}S^{\pi B_i}_N=0.24$,
which is said (by analogy with the nuclear cluster physics) to be
"a total number of pions in the nucleon", shows and approximate
agreement with the "experimental" value $N_{\pi}=0.18\pm0.06$
obtained previously \cite{n} by fitting the old pion knock-out
data \cite{b}.

\section{\textmd Two-pion predictions}

\begin{figure}[t]
\parbox{.45\textwidth}
 {\epsfig{file=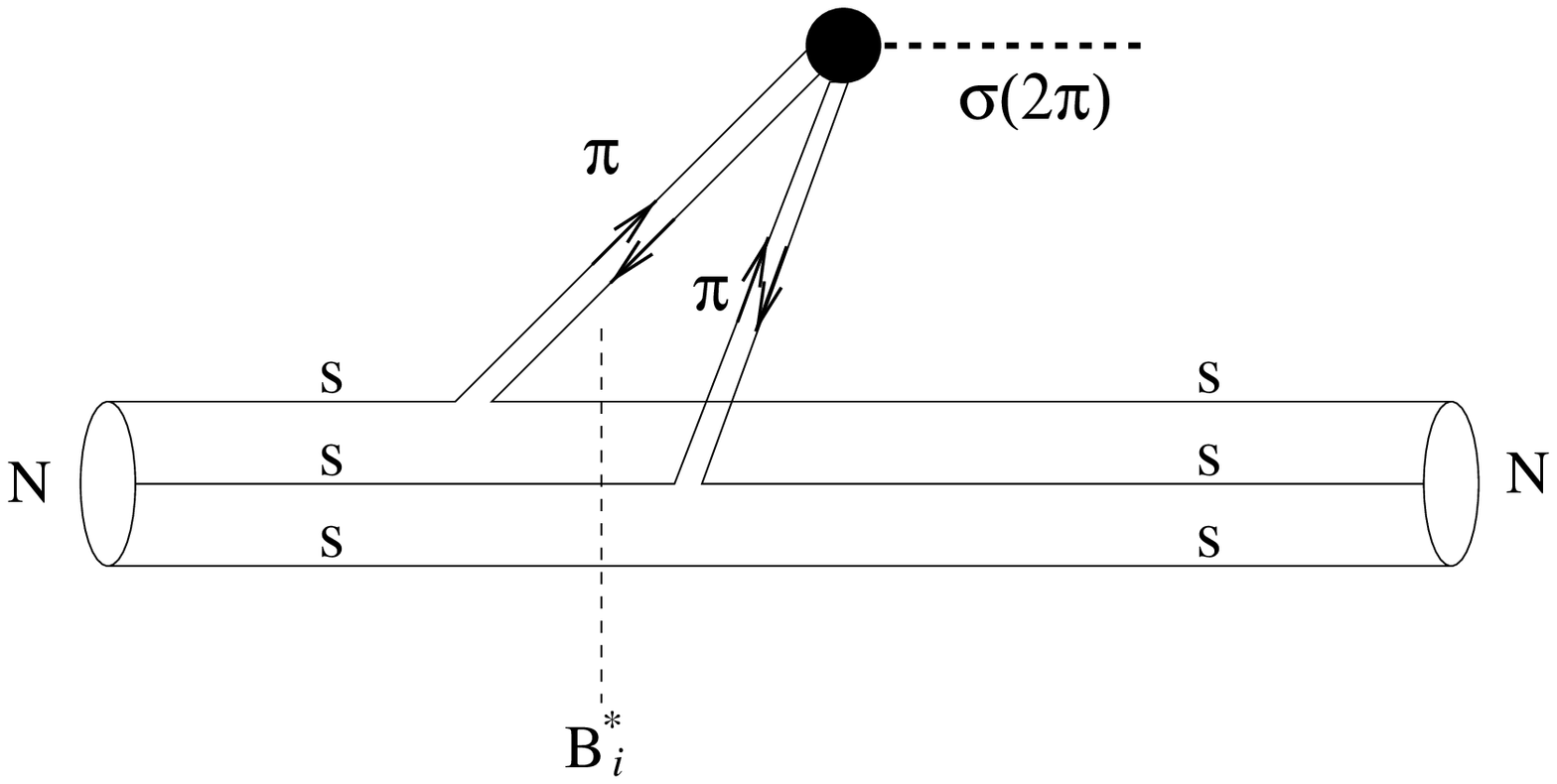,width=0.2\textwidth,angle=-90}
 \caption{\label{f1}Quark diagram of the $N\to\sigma(\pi\pi)+N$
 transition.}}
 \hfill
 \parbox{.5\textwidth}
   {\epsfig{file=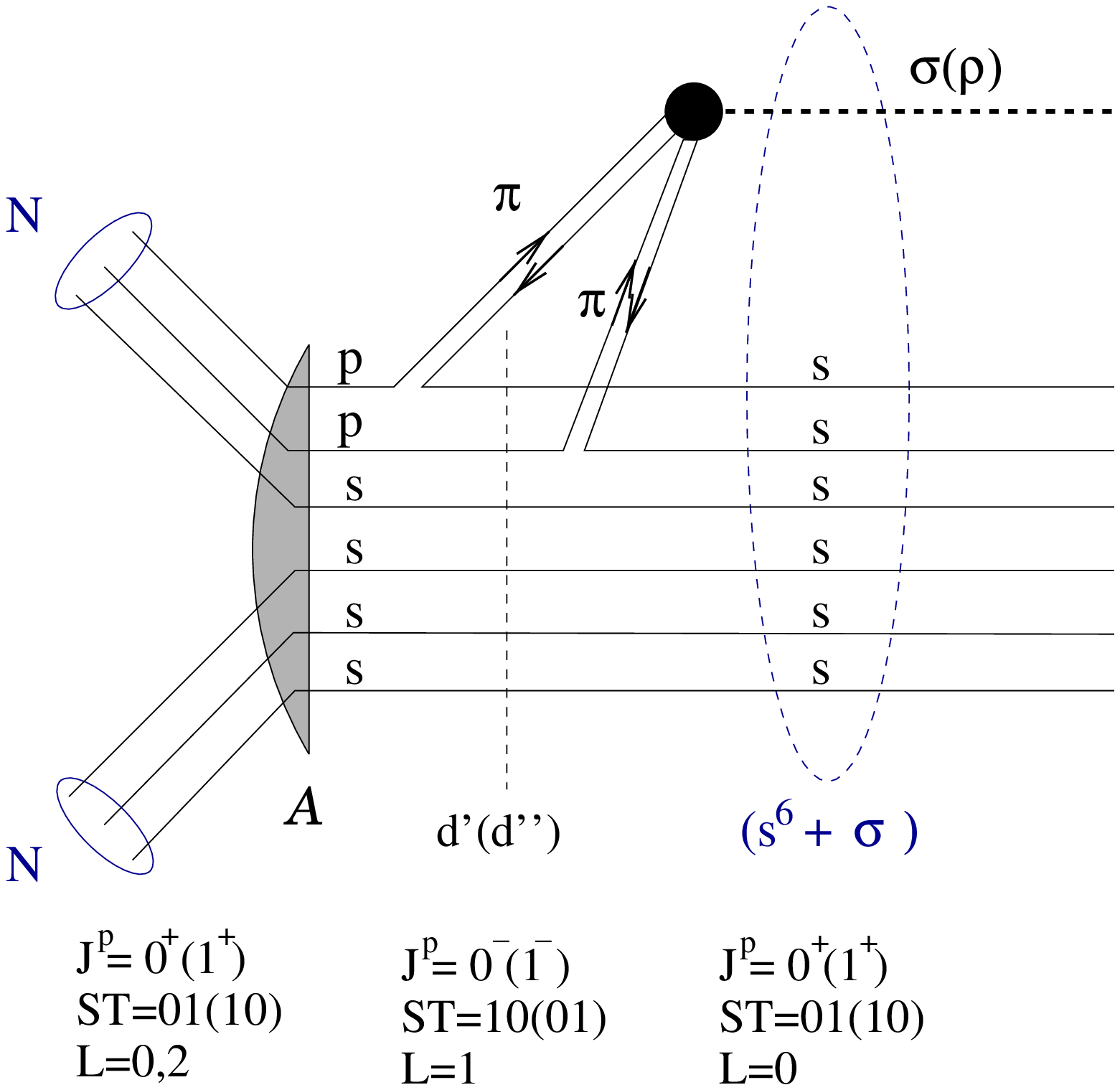,width=0.35\textwidth,angle=-90}
 \caption{\label{f2}The $N+N\to DB$ transition.}}
\end{figure}

In the recent work \cite{k} the $\sigma$-meson formation has been
considered in both the nucleon and the $NN$-system. We start from
the $^3P_0$-model effective $\pi qq$ vertex \cite{k,d,o}
\[
 H^{(3)}_{\alpha}({\b \rho}^{\prime}_2,{\b \rho}_2)=
 \frac{f_{\pi qq}}{m_{\pi}}\,\frac{e^{2i{\bf k}\cdot
 {\b \rho}^{\prime}_2/3}}{\sqrt{2\omega_{\pi}}(2\pi)^{3/2}}\,
 \hat O({\b\rho}^{\prime}_2,{\b\rho}_2)\,\tau^{(3)}_{-\alpha}
 {\b \sigma}^{(3)}
 \cdot\left[\frac{\omega_{\pi}}{2m_q}\left(\frac{2}{i}
 {\nabla}_{\rho_2}+\frac{2}{3}{\bf k}\right)+
 \left(1+\frac{\omega_{\pi}}{6m_q}\right){\bf k}\right],
\]
 and an effective $\sigma\pi\pi$ scalar coupling
 $H_{\pi\pi\sigma}=g_{\pi\pi\sigma}\,\vec\phi_{\pi}
 \cdot\vec\phi_{\pi}\,\phi_{\sigma}$
 ($g_{\pi\pi\sigma}\approx 2-4\,\, GeV$)
with the vertex form factor $F_{\pi\pi\sigma}(q^2)$ of a Gaussian
form $\sim exp(-q^2b_{\sigma}^2/2)$ (here $\hat O$ is a non-local
operator related to the pion $q\bar q$ wave function and
${\b\rho_2=({\bf r}_1+{\bf r}_2)/2-{\bf r}_3}$ is a relative
coordinate). The resulting expression for the $\sigma NN$ coupling
constant $g_{\sigma NN}\equiv g_0\,D_N(0)$ with $g_0=\frac{f_{\pi
 qq}^2}{m_{\pi}^2}\,\frac{g_{\sigma\pi\pi}}{m_q^2b^3}$
depends only on the loop integral $D_N(k)$ for the triangle
diagram in Fig. \ref{f1}. It has been shown that a realistic value
$g_{\sigma NN}\approx$ 5-10 could be only obtained if one takes
into account a {\em coherent superposition} of all the baryon
intermediate states $B_i^*$ in the triangle graph: $N$,
$N^*_{1/2^-}$, $N^*_{3/2^-}$, $N^{**}_{1/2^+}$, $\Delta$,
 $\Delta^*_{1/2^-}$, $\Delta^*_{3/2^-}$, $\Delta^{**}_{3/2^+}$.
In the quark shell model these states correspond to the
configurations up to 2$\hbar\omega$ h.o. $s^3$, $s^2p$, $sp^2$,
and $sp^2-s^22s(2d)$ with all the possible values of spin
$S=1/2,\,3/2$, isospin $T=1/2,\,3/2$, and Young tableau
$[f_X]=[3],\,[21]$. Summation over these states leads to the
following expression:
\[
D_N(k\!=\!0)=\frac{b}{256}\frac{1}{2\pi^2}{\cal P}
\int\limits_0^{\infty}\,q^2dq\,\frac{e^{-\beta^2q^2}f_0(q)\,m_{\sigma}}
{\omega_{\pi}(q/2)\,(m_{\sigma}-2\,\omega_{\pi}(q/2))}\,
\left\{\frac{5}{4}\,\frac{1+q^2b^2/12}
{m_B-m_N+q^2/8m_{B}+\omega_{\pi}(q/2)} \right.
\]
\[ \left. -\,\frac{m_q^2}{\omega^2_{\pi}(q/2)}\,\frac{q^2b^2}{2}
\left[\frac{1}{q^2/8m_N+\omega_{\pi}(q/2)}+
\frac{4}{m_{\Delta}-m_N+q^2/8m_N+\omega_{\pi}(q/2)} \right.
\right.
\]
\[
 \left. \left. +\,\frac{5}{288}\,\frac{q^2b^2\,(1+3q^2b^2/16)}
{m_B-m_N+q^2/8m_N+\omega_{\pi}(q/2)} \right] \right\},\quad
 \beta^2=\frac{b^2}{12}\left(1+\frac{3b_{\sigma}^2}{2b^2}\right),
\]
where the function $f_0$ takes into account all the possible time
orderigs in the triangle diagram \cite{k}
(for simplicity the common mass $m_B=1500$ MeV is used for all the
baryons except $\Delta$ and $N$). However, for realistic values of
the free parameters $b=0.5$ fm, $b_{\sigma}=0.2$ fm and
$g_{\sigma\pi\pi}=4$ GeV we obtain a relatively small value
$g_{\sigma NN}=0.433$. It shows that the high-momentum behaviour
of vertex form factors $N\to B^*+\pi$, $B^*\to N+\pi$ and
$\pi+\pi\to\sigma$ plays an important role in deriving a realistic
value of the coupling constant $g_{\sigma NN}$. The standard quark
shell model with a characteristic (confinement) scale $b\approx$0.5 - 1
fm predicts too soft vertex form factors, and thus some
generalization of the model is needed to incorporate a new scale
parameter of about 0.2-0.3 fm (the inverse value of the
characteristic chiral scale 0.6 - 1 GeV/c). The following
modification of a simple Gaussian form:
$e^{-\beta^2q^2}\,\to\,\frac{2}{3}\,e^{-\beta^2q^2}+
\frac{1}{3}\,e^{-(\beta/Z)^2q^2},\mbox{ with } Z\simeq 2$
 was used to imitate a power-like behaviour
$\sim\left(\frac{\Lambda^2}{q^2+\Lambda^2}\right)$ of the
vertices. At $Z=1.9$ (which corresponds to $\Lambda=0.9$~GeV/c) a
more realistic value $g_{\sigma NN}=3.26$ has been obtained.

\noindent
 The developed model has been used for the two-nucleon
system. A special feature of the above processes in the $NN$
system (see Fig. \ref{f2}) is that a sum over intermediate
dibaryon six-quark states $^2\!B_i^*+\pi$ should be taken into
consideration (hand by hand with the sum $\sum_iB^*_i+\pi$ for
each nucleon). The quantum numbers of intermediate dibaryons are
fixed by the selection rules. In the channel $J^{P}=0^+(ST=01)$
the $^2B ^*$ has quantum numbers of the so-called $d^{\prime}$
dibaryon~\cite{o}
$d^{\prime}=|s^5p[51]_XL\!=\!1,[321]_{CS}ST=10,J^P=0^->$, while in
the "deuteron channel" $J^{P}=1^+(ST=10)$ it has quantum numbers
of the $d^{\prime\prime}$ dibaryon (a counterpart of the
$d^{\prime}$ for $S\rightleftarrows T$ interchange~\cite{k}). The
calculated amplitude for the transition $N+N\,\to\,^2\!B+
\sigma(2\pi)$, where $^2\!B$ is a compact six-quark configuration
$s^6[6]_X$ (a "quark core" of a "dressed bag" state $DB=^2\!B+
\sigma(2\pi)$), leads to a non-diagonal potential $V_{DB,N\!+\!N}$
for two coupled channels ($DB$ and $N\!+\!N$). The corresponding
coupled-channel model of $NN$ interaction proposed in
Ref.~\cite{k} on the basis of the above mechanism leads to a very
reasonable description of the $NN$ scattering at intermediate
energies \cite{k }. The DB admixture in deuteron $P_{DB}$ has been
also calculated \cite{k}: $P_{DB}=3.66$\%. This small value seems
to be very important for a high-precision description of the
deuteron proprties.

\section*{Acknowledgments}The author gratefully acknowledges the
contribution to the talk of all the collaborators of the Moscow
and Moscow-Tuebingen groups: Profs. A. Faessler, V.I. Kukulin,
V.G. Neudatchin, N.P. Yudin, and Drs. V.M. Pomerantsev and L.L.
Sviridova. The work was supported in part by the Deutsche
Forschungsgemeinschaft grant No. Fa-67/20-1 and the Russion
Foundation for Basic Research grants No. 01-02-04015 and
00-02-16117.


\end{document}